\documentclass[aps, prb, preprint, amsmath,amssymb,showpacs,superscriptaddress, longbibliography]{revtex4-1}
\usepackage{graphicx}
\usepackage{ulem}
\usepackage{caption}
\usepackage{color}
\definecolor{green}{RGB}{34,139,34}
\makeatletter

\begin{document}

\title{Unexpected changes in the band structure within AFM1 state of CeBi}

\author{Yevhen Kushnirenko}
\email[]{yevhenk@ameslab.gov}
\affiliation{Division of Materials Science and Engineering, Ames Laboratory, Ames, Iowa 50011, USA}
\affiliation{Department of Physics and Astronomy, Iowa State University, Ames, Iowa 50011, USA}

\author{Brinda Kuthanazhi}
\affiliation{Division of Materials Science and Engineering, Ames Laboratory, Ames, Iowa 50011, USA}
\affiliation{Department of Physics and Astronomy, Iowa State University, Ames, Iowa 50011, USA}

\author{Benjamin Schrunk}
\affiliation{Division of Materials Science and Engineering, Ames Laboratory, Ames, Iowa 50011, USA}

\author{Evan O'Leary}
\affiliation{Division of Materials Science and Engineering, Ames Laboratory, Ames, Iowa 50011, USA}
\affiliation{Department of Physics and Astronomy, Iowa State University, Ames, Iowa 50011, USA}

\author{Andrew Eaton}
\affiliation{Division of Materials Science and Engineering, Ames Laboratory, Ames, Iowa 50011, USA}
\affiliation{Department of Physics and Astronomy, Iowa State University, Ames, Iowa 50011, USA}

\author{Robert-Jan Slager}
\affiliation{TCM Group, Cavendish Laboratory, University of Cambridge, Cambridge CB3 0HE, United Kingdom}

\author{Junyeong Ahn}
\affiliation{Department  of  Physics,  Harvard  University,  Cambridge  MA  02138,  USA}

\author{Lin-Lin Wang}
\affiliation{Division of Materials Science and Engineering, Ames Laboratory, Ames, Iowa 50011, USA}

\author{P. C. Canfield}
\affiliation{Division of Materials Science and Engineering, Ames Laboratory, Ames, Iowa 50011, USA}
\affiliation{Department of Physics and Astronomy, Iowa State University, Ames, Iowa 50011, USA}

\author{Adam Kaminski}
\email[]{kaminski@ameslab.gov}
\affiliation{Division of Materials Science and Engineering, Ames Laboratory, Ames, Iowa 50011, USA}
\affiliation{Department of Physics and Astronomy, Iowa State University, Ames, Iowa 50011, USA}

\maketitle
{\bf We perform angle-resolved photoemission spectroscopy (ARPES) measurements in conjunction with density functional theory (DFT)  calculations to investigate the evolution of the electronic structure of CeBi upon a series of antiferromagnetic (AFM) transitions. We find evidence for a new AFM transition in addition to two previously known from transport studies. We demonstrate the development of an additional Dirac state in the $(+-+-)$ ordered phase and a transformation of unconventional surface-state pairs in the $(++--)$ ordered phase. This revises the phase diagram of this intriguing material, where there are now three distinct AFM states below T$_N$ in zero magnetic field instead of two as it was previously thought.}
\section*{Introduction}
The rare-earth monopnictides family of antiferromagnets has been extensively studied for its magnetism for many years \cite{Tsuchida_1965, tsuchida1967field, nereson_1971, cable1972magnetic, halg1982critical, kasuya1996normal, wiener2000magnetic, pittini1996kerr,yoshikawa2024semimetallic}. Proposals for rare-earth monopnictides being hosts to novel topological states sparked renewed interest in these materials over the last decade \cite{GuoNPJ2017, li2017predicted, DuanCommPhys2018, ZhuPRB2020, zeng2015topological}. These predictions were followed by a series of experimental studies that indeed reported the presence of Dirac states \cite{zeng2016compensated, niu2016presence, lou2017evidence, Kuroda_2018, li2018tunable, SatoCeBi, SakhyaNdSb2022}. It was also shown that these states, in some cases, may develop a substantial gap below Neel temperature (T$_N$) upon establishing long-range antiferromagnetic order \cite{honma2023antiferromagnetic}. 
A recent ARPES study of NdBi \cite{SchrunkNature2022} in the AFM phase discovered the emergence of novel surface states (SS)  and unconventional magnetic band splitting (Kaminski-Canfield splitting) that lead to the formation of spin-textured Fermi arcs in a material that is not a Weyl semimetal. The subsequent studies have shown that such states are also present in several other rare-earth monopnictides \cite{kushnirenko2022rare, kushnirenko2023directional, honma2023unusual, honma2023antiferromagnetic, li2023origin}. An STM study \cite{huang2023hidden} pointed to the presence of a hidden transition between AFM 1q and 2q types of AFM order in NdSb. While the interplay and possible relations to multi-q ordering is subtle~\cite{wang2022multi, li2023origin}, it is evident that this class of materials, due to the experimentally reproducible manifestations, hosts a promising avenue to explore.

CeBi is reported to exhibit two AFM phases in zero magnetic field and low temperature. The transition to the first phase with $(+-+-)$ order occurs at $T =$~25~K, and the transition to the second phase with $(++--)$ order occurs at $T =$~12.5~K (see Fig.~1a) \cite{bartholin1979hydrostatic, rossat1983magnetic, kuthanazhi2022magnetisation}. The full phase diagram of CeBi is shown in Fig.~4f. 
An earlier ARPES study of this material \cite{kushnirenko2022rare} demonstrated the presence of unconventional SS pairs in the AFM1 phase, similar to those in NdBi and NdSb, albeit with much smaller energy splitting. Another ARPES study \cite{SatoCeBi} demonstrated changes in bands near the $\Gamma$-point upon magnetic transition. 

In this study, we use ARPES measurements to investigate the evolution of the electronic structure of CeBi at low temperatures in more detail. Our results demonstrate that in addition to these two well-known transitions, there is an additional one that occurs at $T =$~20~K, although there are no clear signatures in transport nor thermodynamical properties \cite{kuthanazhi2022magnetisation}.

\section*{Results and Discussion}
DFT calculations predict an electronic band structure of CeBi being similar to the electronic structure of other rare-earth monopnictides. The Fermi surface is formed of several hole pockets at the center of the Brillouin zone (BZ) and ellipsoidal electron pockets in each $X$-point of the BZ (Fig.~1b). Similarly to several other rare-earth monopnictides, a band inversion between Bi $6p$ and Ce $5d$ bands occurs along $\Gamma-X$ direction (Fig.~1c). This band inversion should lead to the formation of one SS Dirac cone at the $\overline{\Gamma}$ and two cones at $\overline{M}$-point of the 2D BZ \cite{Kuroda_2018, honma2023antiferromagnetic}. The FS map (Fig.~1e) measured in the paramagnetic (PM) state using ARPES is in good agreement with the DFT calculations (Fig.~1d).

\begin{figure*}[b]
    \includegraphics[width=1\linewidth]{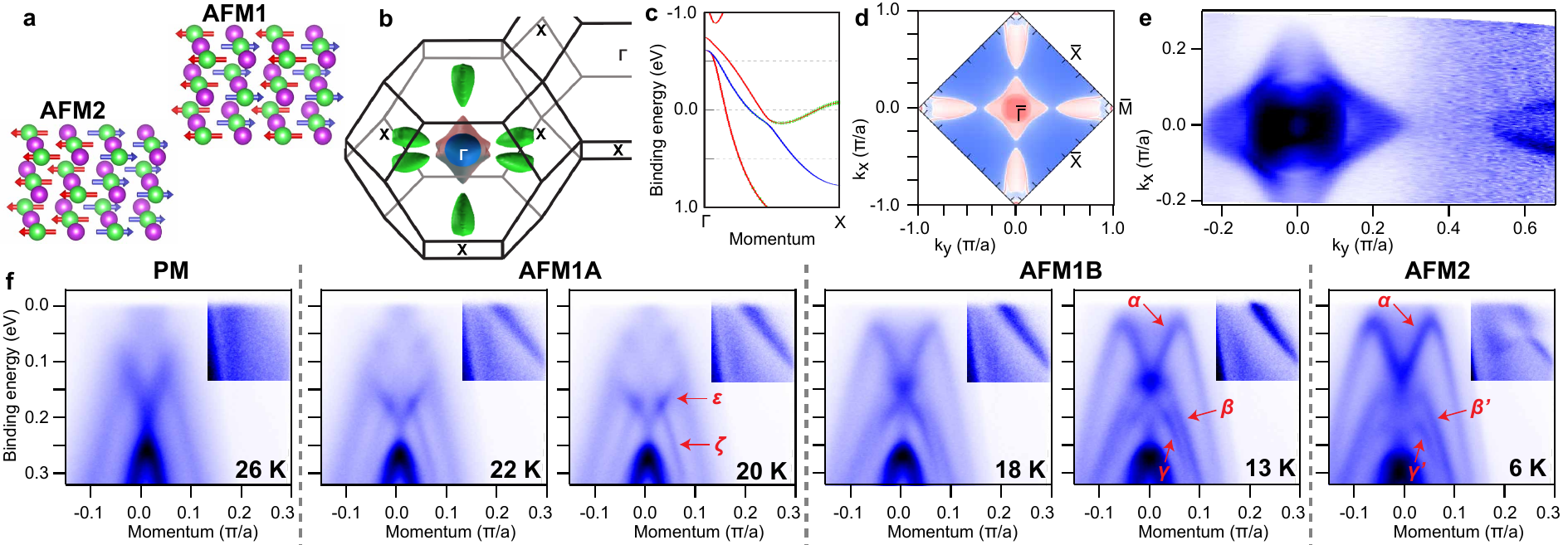}%
    \caption{
    (a) Schematic image of magnetic structure in the AFM1 and AFM2 phases.
    (b) Schematic image of the BZ and the 3D Fermi surface of paramagnetic CeBi.
    (c) DFT calculated electronic structure along $\Gamma-X$ direction.
    (d) DFT calculated 2D projection of bulk Fermi surface.
    (e) Experimental Fermi surface measured in the PM phase.
    (f) ARPES spectra measured along $\overline{\Gamma}-\overline{M}$ direction at different temperatures. The insets in the top-right corner demonstrate corresponding parts of the spectra with increased intensity. The second derivative images and the spectra measured at T = 10, 16, 24K can be found in the Supplementary information \cite{SM}.
    }
\end{figure*}

In Fig.~1f, we show the temperature evolution of the measured band dispersion near the center of the BZ. In the PM state, for T=26~K, band dispersion at $\overline{\Gamma}$ consists of several hole-like bulk bands that appear very broad due to projection along $k_z$ direction. There are no states present in the square located in the upper right corner, where the intensity is enhanced by a factor of 20. Upon cooling to 22~K, the sample undergoes the first AFM transition into the AFM1 phase, which leads to the appearance of a surface state (linear dispersion crossing $E_f$  in the square of enhanced intensity).  The presence of these states was already reported in a recent study \cite{ kushnirenko2022rare}.  The splitting between two SS in this material is quite small, thus they appear in the spectra as a single feature. In addition, two more surface states emerge near $\overline{\Gamma}$, which are marked as $\epsilon$ and $\zeta$ bands in Fig.~1f that are present in data measured at 20~K and 22~K.

There are several differences between higher temperature AFM1 phase (T=20~K and 22~K) and  AFM2 phase at 6~K.  The $\epsilon$ and $\zeta$ bands vanish, while three new bands marked as $\alpha$, $\beta'$, and $\gamma'$ appear. In addition, the sharp surface state in the square of enhanced intensity develops an energy gap at $\sim$40 meV. The SS, which was essentially a single continuous band at T=22~K, is now split into two segments separated by an area of low intensity in the AFM2 phase at T=6~K, which will be discussed later in more detail. 

While expected to be similar to the 20~K and 22~K spectra, the spectra measured at T~=~13 and 18~K are different from them as well as from the AFM2 spectrum.
In the 13~K spectrum, the surface states in the top-right corner are no longer gapped in contrast to ones in the 6~K spectrum, and this feature looks like the one in the 20~K spectrum. Also, despite the general similarity of bands near the $\overline{\Gamma}$-point in 13~K and 6~K spectra, we observe additional intensity near the bottom of the $\alpha$ band in 13~K spectrum. These changes are better seen from the comparison of momentum distribution curves (MDC) in Fig.~2a. While the peaks at $k_x > 0.1 \pi/a$ (associated with the outer part of the $\alpha$ band) do not change upon the transition, the central peak is more intense at 13 K. The corresponding peak on the zero momentum energy distribution curve (EDC) at 13 K is higher and narrower as compared to the 6~K curve (Fig.~2b). This happens because in the AFM1 phase, the top part of the $\beta$ band overlaps with the bottom of the $\alpha$ band (see also 2nd derivative images in the Supplementary information \cite{SM}), while in the AFM2 phase, the $\beta'$ band demonstrates a back-folding and does not cross the $\alpha$ band.
At $T =$~20~K, a fairly dramatic change occurs in the electronic structure. The $\alpha$, $\beta$, and $\gamma$ bands disappear, and instead of them, we observe a V-shaped distension with a bottom at $E_B =$ 205 meV (marked as $\epsilon$) and a weak Dirac-like feature with a crossing at $E_B =$ 125 meV (marked as $\zeta$). This new transition that occurs at 20K was not observed before with other techniques.

\begin{figure}[b]
    \includegraphics[width=0.5\linewidth]{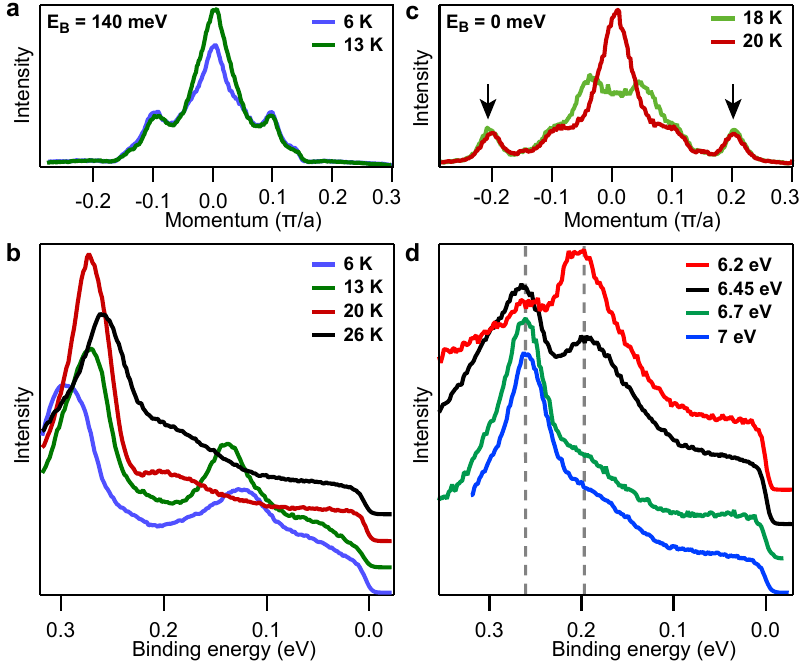}%
    \caption{
    (a) MDCs obtained at $E_B =$ 140 meV from 6 K and 13 K spectra in Fig.~1f.
    (b) Zero-momentum EDCs obtained from spectra in Fig.~1f.
    (c) MDCs obtained at $E_B =$  0 meV, from 18 K and 20 K spectra in Fig.~1f.
    (d) EDCs obtained at zero momentum from the spectra in Fig.~3a.
    }
\end{figure}

One can assume that two different band structures in the AFM1 phase are associated with the presence of domains with different directions of magnetic ordering. Indeed, the presence of domains was shown in two other rare-earth monopnictides: NdBi \cite{honma2023antiferromagnetic} and NdSb \cite{kushnirenko2023directional,honma2023unusual}. However, in these studies, the domains were stationary and did not show a rapid change in the ordering direction. Also, the transition at 20 K in CeBi was observed in multiple samples and was reproducible in the second cycle of cooling down and heating the sample (see Supplementary information\cite{SM}). Finally, the SS associated with the (100) domain (magnetic moments oriented parallel to the sample surface) is present below and above 20 K. Even more, the comparison of the Fermi level MDCs obtained from 18K and 20K spectra in Fig.~2c demonstrates a similar intensity of the peaks associated with the SS pairs (marked with arrows). The small difference is explained by a gradual decrease in the SS intensity with temperature reported before in several rare-earth monopnictides \cite{SchrunkNature2022, kushnirenko2022rare}. All this proves that an additional phase transition occurs at 20 K, and observed effects are not accidental switching of different domains. The further temperature increase does not qualitatively change the electronic structure until the transition to the PM phase at $T =$~25~K. We, therefore, call the phase which exists from $T =$~12.5~K to 20K AFM1B, and the phase which exists from $T =$~20~K to 25~K AFM1A.

To better understand the electronic structure, we measured sets of spectra for the PM, AFM1A, and AFM1B phases using different photon energies. The results are shown in Fig.~3. The plots (a) and (b) demonstrate the spectra measured in the PM phase, and the corresponding second derivative plots. All four spectra demonstrate a Dirac-like feature at $\overline{\Gamma}$-point. This Dirac-like feature was observed in several other rare-earth monopnictides such as name LaBi, PrBi, SmBi, GbBi, NdBi \cite{wu2016asymmetric, niu2016presence, li2018tunable, SakhyaNdSb2022} and is usually associated with a SS Dirac cone predicted to be there as a result of the band inversion along $\Gamma-X$ direction in the 3D BZ \cite{Kuroda_2018, honma2023antiferromagnetic}. However, when tuning the photon energy, we observe changes in the relation of the intensities of the lower and the upper parts of the cone. Also, the zero-momentum EDCs obtained from these spectra (see. Fig.~2d) demonstrate a 60 meV gap in the cone and a small shift with photon energy. In 6.7 and 7 eV EDCs spectra, the upper half of the cone has weak intensity. Because of that, its bottom appears in the corresponding EDCs as a shoulder on the right side of the peak associated with the top of the lower half of the cone. A similar gap was also observed in a sibling compound LaBi \cite{wu2016asymmetric}. The observed band shift and the presence of the gap indicate that the surface states are hybridized with the neighboring bulk states \cite{li2018tunable}. Although, the shift can be a result of small misalignment, since the 6.45~eV spectrum was measured from a different sample\cite{SM}.

\begin{figure}[b]
    \includegraphics[width=0.5\linewidth]{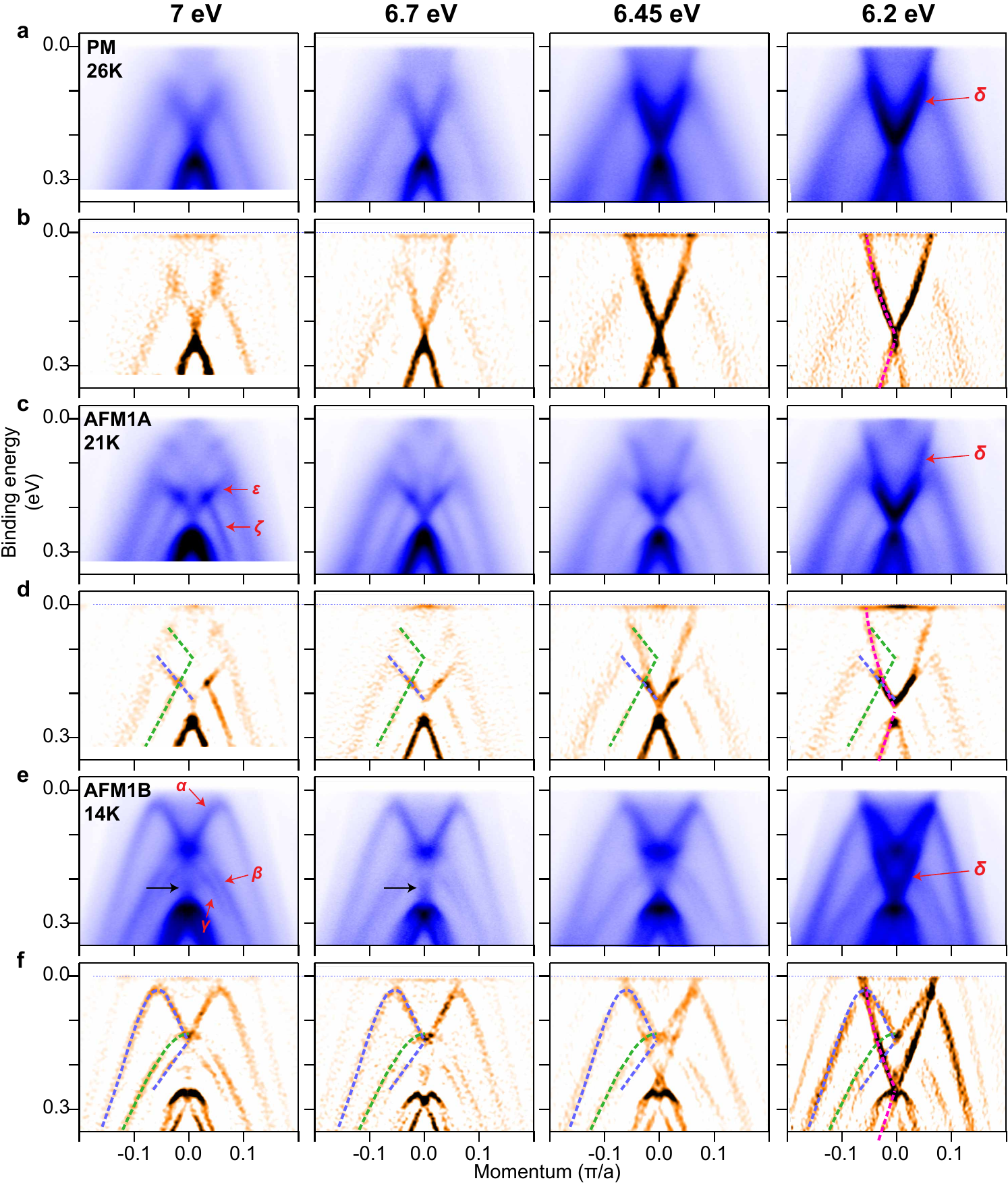}%
    \caption{Photon energy dependence.
    (a, c, e) ARPES spectra measured along $\overline{\Gamma}-\overline{M}$ using different photon energies at 26, 21, and 14 K, respectively.
    (b, d, f) corresponding second derivative images.
    $\epsilon$ and $\zeta$ bands in the AFM1A phase and $\alpha$, $\beta$, $\gamma$ bands in the AFM1B phase do not shift with photon energy indicating their SS nature.}
\end{figure}

All spectra measured in the AFM1A phase (Fig. 3c) show the presence of $\epsilon$ and $\zeta$ bands, and all spectra measured in the AFM1B phase (Fig. 3e) have $\alpha$, $\beta$, and $\gamma$ bands present. Furthermore, the shape of these bands does not change with photon energy. To demonstrate this, we added guides for an eye, which are identical for all spectra. This can indicate that all these bands are due to surface states. Since the measurements were performed in a relatively narrow range of photon energies, we cannot completely rule out that some of these bands may be formed by bulk states with week $k_z$ dispersion. Nevertheless, we consider this scenario unlikely because such states exist in systems with weak interlayer coupling, which is not expected to be the case in CeBi with a cubic rock salt crystal structure. These data also prove that $\alpha$ and $\gamma$ features are actually parts of a single Dirac cone. The spectra measured using 6.2 and 6.45 eV photons in AFM1A and AFM1B phases demonstrate one more feature (marked as $\delta$). From the comparison of these spectra with the spectrum measured in the PM phase, one can see that the $\delta$ feature is present in the upper half of the cone in the PM phase. Except for a small energy shift, the shapes of these bands are identical (as shown with a pink guide for an eye). Some hints of this feature can also be observed in 6.7 and 7 eV spectra (marked with black arrows in Fig.~3e)

\begin{figure}[b]
    \includegraphics[width=0.5\linewidth]{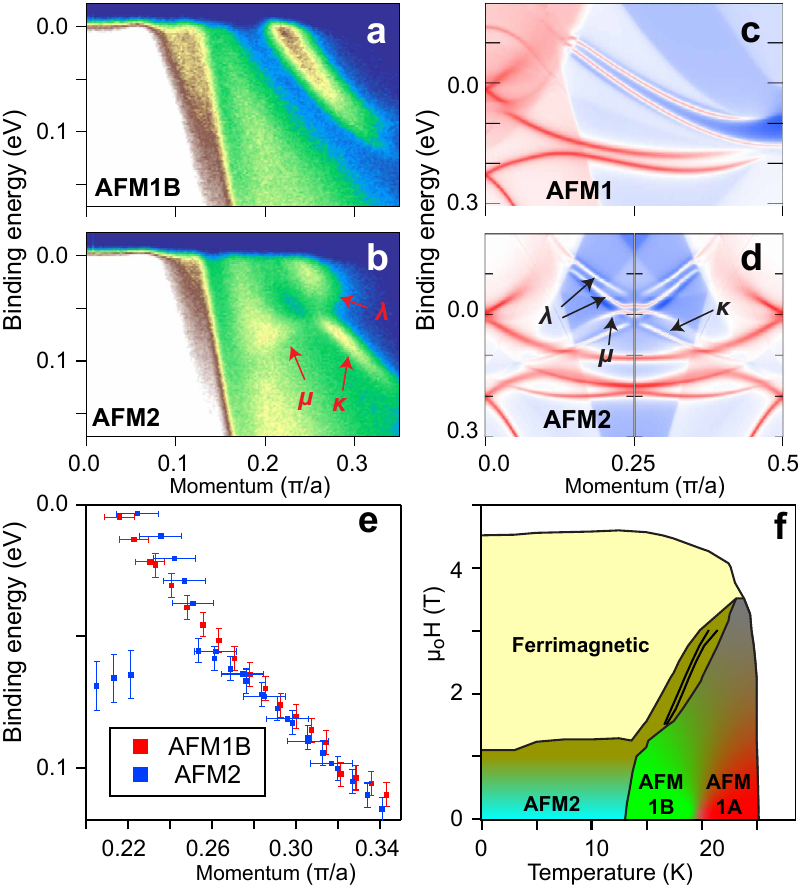}%
    \caption{
    (a-b) ARPES spectra measured along the $\overline{\Gamma}-\overline{M}$ direction in the AFM1B and AFM2 phases, respectively.
    (c-d) DFT calculated band dispersions along the $\overline{\Gamma}-\overline{M}$ direction for AFM1 and AFM2 phases, respectively.
    (e) The band dispersions extracted from the spectra in (a) and (b).
    (f) Phase diagram of CeBi obtained from the magnetic measurements \cite{kuthanazhi2022magnetisation} with added transition between AFM1A and AFM1B.}
\end{figure}

We now discuss the transformation of the unconventional SS  in the AFM2 phase. In Fig.~4a and b, we compare spectra measured along the $\overline{\Gamma}-\overline{M}$ direction with higher statistics in the AFM1B and AFM2 phases. The observation of unconventional SS  in the AFM1A phase is in agreement with the DFT calculations (Fig. 4c). Upon the transition to AFM2 phase, the DFT calculations predict the band structure to be folded one more time \cite{matt2020}. This folding results in opening a gap in both band dispersions of the unconventional SS pair (Fig. 4d). Our ARPES data measured in the AFM2 phase also demonstrates the gap opening in the SS. However, not all SS bands predicted by the calculations can be seen in the experimental spectrum. We can identify three features in this spectrum: $\kappa$, $\lambda$, and $\mu$.

For further analysis, we extract the shape of the SS band from Fig.~4b and the lower SS bands from Fig.~4a by fitting the experimental data. The results are shown in Fig.~4e. The $\kappa$ band in Fig.~4b is sharp and does not show any signs of doubling. Also, its shape and position match the shape and position of the lower SS band in the AFM1A state, as can be seen in Fig.~4e. Thus, we can associate it with the lowest SS band in the second quarter of the BZ. The $\lambda$ feature is broader than the $\kappa$ band. Its broadness is comparable with the broadness of the feature formed by two closely located SS band in the AFM1 phase (at the corresponding energy range). Because of this, we associate the $\lambda$ feature with the superposition of the two upper SS bands in the first quarter of the BZ. The $\mu$ feature has a smaller slope than the one of $\kappa$ band. Thus, we can associate it with the second-lowest SS band in the first quarter of the BZ.
This good correspondence with the DFT calculations indicates that these SS in the AFM2 phase are indeed related to the unconventional SS pair in the first antiferromagnetic phase.

Despite the good correspondence of these surface states with the DFT calculations for the case of order with magnetic moments oriented parallel to the sample surface, these DFT calculations reproduce neither the additional states near the $\Gamma$-point observed in the AFM1A state nor those observed in the AFM1B state. One may attempt to associate an additional Dirac cone at 130~meV formed by $\alpha$ and $\gamma$ dispersions in the AFM1B spectra with one of two Dirac cones predicted to be folded from the $M$-point to the $\Gamma$-point (located at 20 and 220 meV in Fig.~4c and Fig.~S3\cite{SM}). We can exclude this interpretation since, as was shown in an earlier ARPES study \cite{li2018tunable} and our He-lamp ARPES measurements \cite{SM}, the Dirac cones at $M$-point are located considerably lower (at 240 and 400~meV).

\section*{Conclusions}
In conclusion, our ARPES measurements demonstrate the presence of a new phase transition in CeBi. The AFM1A and AFM1B phases, between which the new transition occurs, have displayed significantly different band dispersions, and this transition is marked by the appearance of additional Dirac cones in the AFM1B phase. The nature of this transition is not known. We speculate that it could be due to a change from 1q to 2q antiferromagnetic order, similar to the transition recently reported in a sibling compound NdSb by the STM study \cite{huang2023hidden}. Indeed, the unconventional SS  are predicted to be present in rare-earth monopnictides in the cases of 1q as well as in the cases of 2q ordering~\cite{wang2022multi, li2023origin, wang2023sdh}. Such a proposal would be in reasonable agreement with our experimental results. Also, it is plausible that this transition happens only at the surface. The discrepancy between the magnetic ordering at the surface and in bulk was reported in ferrimagnetically ordered CeBi by STM \cite{matt2020}. However, the STM study is focused on the Ferrimagnetic phase, which exists in the external magnetic field (see phase diagram in Fig.~4f), and the AFM2 phase, while it does not present data for the AFM1 phase.
Establishing the microscopic origin of this transition requires further studies using other experimental techniques, such as neutron scattering or STM with a magnetic tip and magneto-optical polar Kerr.

\bibliography{ndBi_arcs}

\section*{Methods}
Single crystals of CeBi were grown out of In flux. The elements with an initial stoichiometry of Ce$_4$Bi$_4$In$_{96}$ were put into a fritted alumina crucible\cite{Canfield2016Use} and sealed in fused silica tube under partial pressure of argon. The prepared ampules were heated up to 1100$^\circ$~C over 4 hours and held there for 5 hours. This was followed by a slow cooling to the decanting temperature over 100 hours and decanting of the excess flux using a centrifuge.\cite{Canfield_2019}. The decanting temperatures were 850$^\circ$~C. The cubic crystals obtained were stored and handled in a glovebox under Nitrogen atmosphere. 

ARPES data was collected using vacuum ultraviolet (VUV) laser ARPES spectrometer that consists of a Scienta DA30 electron analyzer, picosecond Ti:Sapphire oscillator and fourth-harmonic generator \cite{jiang2014tunable}. Data from the laser based ARPES were collected with 6.7~eV photon energy. Angular resolution was set at $\sim$ 0.1$^{\circ}$ and 1$^{\circ}$, along and perpendicular to the direction of the analyzer slit, respectively, and the energy resolution was set at 2 meV. The VUV laser beam was set to vertical polarization. The diameter of the photon beam on the sample was $\sim 15\,\mu$m.
Samples were cleaved \textit{in-situ} along (001) plane, usually producing very flat, mirror-like surfaces. The measurements were performed at a base pressure lower than 2$\times$10$^{-11}$ Torr. Results were reproduced using several different single crystals of material, and extensive temperature cycling.

Density functional theory \cite{Hohenberg1964Inhomogeneous, Kohn1965Self} calculations with spin-orbit coupling (SOC) were performed for CeBi with the PBE\cite{Perdew1996Generalized} exchange-correlation functional using a plane-wave basis set and projector augmented wave method\cite{Blochl1994Projector}, as implemented in the Vienna Ab-initio Simulation Package (VASP)\cite{Kresse1996Efficiency, Kresse1996Efficient}. In the DFT calculations, for example, with AFM1, we used a kinetic energy cutoff of 300~eV, $\Gamma$-centered Monkhorst-Pack\cite{Monkhorst1976Special} (11$\times$11$\times$8) $k$-point mesh with the experimental lattice constant\cite{leger1993chalcogenides} of 6.55~\AA, and a Gaussian smearing of 0.05~eV. To account for the strongly localized Ce~4$f$ orbitals, an onsite Hubbard-like\cite{liechtenstein1995density} U~=~7.0~eV and J~=~0.9~eV have been used. Our DFT+U+SOC calculation gives a spin moment of 0.1~$\mu_B$ and an orbital moment of 2.2~$\mu_B$ in the same direction, resulting in a total magnetic moment of 2.3~$\mu_B$ on Ce, agreeing well with the experimental data\cite{nereson_1971} of 2.0$\pm$0.1~$\mu_B$. Using maximally localized Wannier functions\cite{Marzari1997Maximally, Souza2001Maximally}, tight-binding models were constructed to reproduce closely the band structure including SOC within EF$\pm$1eV with Ce~$s-d-f$ and Bi~$p$ orbitals. The surface spectral function and 2D Fermi surface (FS) were calculated with the surface Green’s function methods\cite{Sancho1984Quick, Sancho1985Highly} as implemented in WannierTools\cite{wu2018wanniertools}.

\section*{Data Availability}
All data are available from the corresponding author upon reasonable request.

\begin{acknowledgments}
This work was supported by the U.S. Department of Energy, Office of Basic Energy Sciences, Division of Materials Science and Engineering. Ames Laboratory is operated for the U.S. Department of Energy by Iowa State University under Contract No. DE-AC02-07CH11358. R.-J.S. acknowledges funding from a New Investigator Award, EPSRC grant EP/W00187X/1, a EPSRC ERC underwrite grant  EP/X025829/1 as well as Trinity College, Cambridge.
\end{acknowledgments}

\section*{Author contributions}
B.K. and P.C.C grew and characterized the samples. Y.K., B.S, A.K., E.O'L and A.E. performed ARPES measurements and support. Y.K. performed analysis of ARPES data. L.-L.W. performed DFT calculations. R.-J.S. and J.A. provided theoretical analysis. The manuscript was drafted by Y.K. and A.K. All authors discussed and commented on the manuscript.

\section*{Competing interests} The authors declare no competing interests.

\end{document}